\documentclass[12pt]{article}
\usepackage{amsmath}
\usepackage{amssymb}
\usepackage{amsfonts}
\usepackage{graphics}

\title{ Rational extensions of the trigonometric Darboux-P\"{o}schl-Teller
potential based on para-Jacobi polynomials}
\author{B.\ Bagchi{\small \textsl{$^{1}$}}, Y.\ Grandati{\small \textsl{$
^{2}$}} and C. Quesne{\small \textsl{$^{3}$}} \\
{\small \textsl{$^{1}$Department of Applied Mathematics, University of
Calcutta,}}\\
{\small \textsl{92 Acharya Prafulla Chandra Road, Kolkata, India}}\\
{\small \textsl{bbagchi123@gmail.com}}\\
{\small \textsl{$^{2}$Equipe BioPhysStat, LCP A2MC, Universit\'{e} de
Lorraine-Site de Metz,}}\\
{\small \textsl{1 bvd D. F. Arago, F-57070, Metz, France}}\\
{\small \textsl{grandati@univ-lorraine.fr}}\\
{\small \textsl{$^{3}$Physique Nucl\'{e}aire Th\'{e}orique et Physique Math%
\'{e}matique, Universit\'{e} Libre de Bruxelles,}} \\
{\small \textsl{Campus de la Plaine CP229, Boulevard~du Triomphe, B-1050
Brussels, Belgium}}\\
{\small \textsl{cquesne@ulb.ac.be}}}
\date{ }
\begin{document}
\baselineskip=22pt plus 1pt minus 1pt
\maketitle

\begin{abstract}

The possibility for the Jacobi equation to admit in some cases general solutions that are polynomials has been recently highlighted by Calogero and Yi, who termed them para-Jacobi polynomials. Such polynomials are used here to build seed functions of a Darboux-B\"acklund transformation for the trigonometric Darboux-P\"oschl-Teller potential. As a result, one-step regular rational extensions of the latter depending both on an integer index $n$ and on a continuously varying parameter $\lambda$ are constructed. For each $n$ value, the eigenstates of these extended potentials are associated with a novel family of $\lambda$-dependent polynomials, which are orthogonal on $\left] -1,1\right[ $.
 
\end{abstract}

\baselineskip=22pt plus 1pt minus 1pt 

\noindent Running title: Rational extensions


%
%
\newpage

\section{INTRODUCTION}

The construction of new solvable systems is an interesting problem by itself in
mathematical physics. In the quantum framework, the techniques of Supersymmetric
Quantum Mechanics (SUSY QM) \cite{cooper, cooper-bis, bijan} have proved to be one of the
simplest and efficient tools to achieve this goal. From the analytical point
of view, these basically correspond to a single or multi-step Darboux
transformations \cite{darboux, crum}. Applying such transformations to a
class of known solvable systems that are guided by primary translationally shape invariant
potentials (TSIP) \cite{cooper, cooper-bis, bijan, gendenshtein, grandatiberard}, we can
generate chains of new potentials, called extensions, which are
(quasi)isospectral to the original potential and whose eigenfunctions are
entirely determined in an explicit way.\par
%
%
Of particular interest are the rational extended potentials, i.e., extensions
which are rational in an appropriate variable. It appears that those
obtained from the three confining TSIP (namely, the 
harmonic oscillator, isotonic
oscillator and trigonometric Darboux-P\"{o}schl-Teller potentials) have
eigenfunctions which coincide (up to a gauge factor) with the exceptional
orthogonal polynomials recently discovered by G\'omez-Ullate, Kamran, and Milson
\cite{gomez3, gomez4, quesne1, cq09, odake09}. During the last five years, this topic
has been the subject of an intense research activity (see \cite{GGM1} and
references therein).\par
%
%
In general, for a given primary TSIP, the possible rational extensions are
characterized by one or two multi-indices, which define in a unique way the
corresponding extended potentials \cite{gomez5, odake3, cq11b, grandati11b, gq7}. Then,
until now, the rationality condition appears quite restrictive, by excluding
the possibility to generate families of potentials depending on one or
several continuous parameters, which is a very interesting feature of SUSY QM
\cite{keung, wang}. The one ($n$-)parameter isospectral families are in
particular closely connected to one (multi-)soliton solutions of some nonlinear evolution equations.
\par
%
%
In this paper, we show that for the trigonometric Darboux-P\"{o}schl-Teller (TDPT) potential with some specific underlying parameters, it is nevertheless possible to generate one-parameter families of rational
extensions via one-step state-adding Darboux transformations. To achieve
this goal we use seed functions based on para-Jacobi polynomials \cite{calogero}. Although noticed by Szeg\"{o} in his classical treatise \cite{szego}, the possibility for the Jacobi equation to admit in some cases a
general solution that is a polynomial has not been considered in the other
standard treatises on orthogonal polynomials. It has been highlighted only
recently by Calogero and Yi \cite{calogero}, who introduced the so-called
para-Jacobi polynomials. This result has been now extended by Calogero, who
has considered in particular a more general second-order linear differential
equation, featuring an arbitrary number of free parameters, for which all the
solutions are polynomials \cite{calogero2}, and has also defined the concept
of generalized hypergeometric polynomials \cite{calogero3}.\par
%
%
The paper is organized as follows. We start by recalling the essential features
of Darboux transformations and of the trigonometric Darboux-P\"{o}%
schl-Teller potential (TDPT) in Secs.~II and III, respectively. In Sec.~IV, we introduce the
para-Jacobi polynomials and specify the domain of values of the free
parameter for which they are nodeless on the considered interval. In Sec.~V, we then use the corresponding eigenfunctions as seed functions for
state-adding Darboux-B\"acklund transformations to build a set of extensions of the TDPT potentials, which
are regular, rational, and dependent on one continuous parameter. In
addition, we determine explicitly the eigenstates of these extended
potentials and give plots for several values of the parameter in the first non-trivial case of the extension. Finally, Sec.~VI contains the conclusion.\par
%
%
\section{DARBOUX TRANSFORMATIONS AND EXTENDED SOLVABLE POTENTIALS}

A formal eigenfunction $\psi _{E}(x)$ for the eigenvalue $E$ of the one-dimensional Hamiltonian 
$\widehat{H}=-d^{2}/dx^{2}+V(x),\ x\in I\subset 
\mathbb{R}$, is a solution of the Schr\"{o}dinger equation
\begin{equation}
\psi _{E}^{\prime \prime }(x)+\left( E-V(x)\right) \psi _{E}(x)=0.
\label{EdS}
\end{equation}
In the following, we simply call eigenfunctions such formal eigenfunctions,
employing eigenstates for the eigenfunctions satisfying given Dirichlet
boundary conditions at the limits of the definition interval $I$. With such
Dirichlet boundary conditions on $I$, $\widehat{H}$ is supposed to admit a
discrete spectrum of energies and eigenstates $\left( E_{n},\psi _{n}\right)
_{n\in \left\{ 0,...,n_{\max }\right\} \mathbb{\subseteq N}}$ where, without
loss of generality, we can always take the ground level of $\widehat{H}$ at
zero ($E_{0}=0$). We suppose also that in the considered domain of
eigenvalues, the eigenfunctions can be indexed by a real spectral parameter $%
\mu $ ($E\rightarrow E_{\mu },$ $\psi _{E}\rightarrow \psi _{\mu }(x)$)
which takes integer values $\mu =n\in 
\mathbb{N}
$ when $E_{n}$ is a bound-state energy.\par
%
%
Starting from an eigenfunction $\psi _{\nu }(x)$ of $\widehat{H}$, we define
the first-order operator $\widehat{A}\left( w_{\nu }\right) $ by
\begin{equation}
\widehat{A}\left( w_{\nu }\right) =d/dx+w_{\nu }(x),  \label{opA}
\end{equation}
where $w_{\nu }(x)=-\psi _{\nu }^{\prime }(x)/\psi _{\nu }(x)$. Then, if $%
W\left( y_{1},...,y_{m}\mid x\right) $ denotes the Wronskian of the
functions $y_{1},...,y_{m}$ \cite{muir},
\begin{equation}
W\left( y_{1},...,y_{m}\mid x\right) =\left\vert 
  \begin{array}{ccc}
  y_{1}\left( x\right)  & ... & y_{m}\left( x\right)  \\ 
  ... &  & ... \\ 
  y_{1}^{\left( m-1\right) }\left( x\right)  & ... & y_{m}^{\left( m-1\right)
     }\left( x\right) 
  \end{array}
\right\vert ,  \label{wronskien}
\end{equation}
for $\mu \neq \nu $, the function defined via the so-called Darboux-Crum
formula 
\begin{equation}
  \psi _{\mu }^{\left( \nu \right) }=\widehat{A}\left( w_{\nu }\right) \psi
  _{\mu }(x)=\frac{W\left( \psi _{\nu },\psi _{\mu }\mid x\right) }{\psi
  _{\nu }(x)}  \label{DC}
\end{equation}
is a solution of the Schr\"{o}dinger equation
\begin{equation}
  \psi _{\mu }^{\left( \nu \right) \prime \prime }(x)+\left( E_{\mu
  }-V^{\left( \nu \right) }(x)\right) \psi _{\mu }^{\left( \nu \right) }(x)=0,
\end{equation}
with the same energy $E_{\mu }$ as in Eq.~(\ref{EdS}), but with a modified
potential
\begin{equation}
  V^{\left( \nu \right) }(x)=V(x)+2w_{\nu }^{\prime }(x).  \label{pottrans}
\end{equation}
\par
%
%
We call $V^{\left( \nu \right) }(x)$ an extension of $V(x)$ and the
correspondence
\begin{equation}
  \left(\begin{array}{c}
  V(x) \\
  \psi _{\mu }(x)
  \end{array}
    \right) \overset{\widehat{A}\left( w_{\nu }\right) }{\rightarrow }
  \left(\begin{array}{c}
  V^{\left( \nu \right) }(x) \\ 
  \psi _{\mu }^{\left( \nu \right) }(x)
  \end{array}\right)
\end{equation}
is called a Darboux-B\"{a}cklund Transformation (DBT). The eigenfunction $
\psi _{\nu }$ is the seed function of the DBT $\widehat{A}(w_{\nu })$.\par
%
%
Note that $\widehat{A}(w_{\nu })$ annihilates $\psi _{\nu }$ and,
consequently, Eq.~(\ref{DC}) is not valid for $\mu =\nu $.
Nevertheless, we can readily verify that $1/\psi _{\nu }(x)$ is an
eigenfunction of $V^{\left( \nu \right) }(x)$ for the eigenvalue $E_{\nu }$.
Consequently, by extension, we define the ``image'' by $\widehat{A}(w_{\nu })$ of the
seed eigenfunction $\psi _{\nu }$ itself as
\begin{equation}
  \psi _{\nu }^{\left( \nu \right) }(x)\sim 1/\psi _{\nu }(x).  \label{psinunu}
\end{equation}
\par
%
%
In general, the transformed potential $V^{\left( \nu \right) }(x)$ is
singular at the nodes of $\psi _{\nu }(x)$ and for integer values $n$ of $\nu $, 
$V^{\left( n\right) }$ is regular only when $n=0$, that is, when the seed
function is the ground state of $\widehat{H}$, which corresponds exactly to
the usual SUSY partnership in quantum mechanics \cite{cooper,cooper-bis,bijan}.\par
%
%
We can nevertheless envisage to use as seed function an eigenfunction
associated to an eigenvalue in the disconjugacy sector of Eq.~(\ref{EdS}),
here a negative eigenvalue \cite{grandati11b}. Indeed, in this sector,
every solution of this equation has at most one (simple) zero on $I$ \cite
{hartman,coppel} and, moreover, the existence of nodeless solutions of this
equation is ensured.\par
%
%
To prove that a given solution $\psi _{\nu }(x)$ belongs to this category,
it is sufficient to determine its signs at the boundaries of $I$. If they
are identical, then $\psi _{\nu }$ is nodeless and if they are opposite, then 
$\psi _{\nu }$ presents a unique zero on $I$. In the first case, $V^{\left(
\nu \right) }(x)$ constitutes then a regular (quasi)isospectral extension
of $V(x)$.\par
%
%
\fussy
\section{TRIGONOMETRIC DARBOUX-P\"OSCHL-TELLER (TDPT) POTENTIAL}
\setcounter{equation}{0}

The trigonometric Darboux-P\"{o}schl-Teller (TDPT) potential (with zero
ground-state energy) is defined on$\ x\in \left] 0,\pi /2\right[ $ by
\begin{equation}
  V(x;\alpha ,\beta )=\frac{(\alpha +1/2)(\alpha -1/2)}{\sin ^{2}x}+\frac{
  (\beta +1/2)(\beta -1/2)}{\cos ^{2}x}-(\alpha +\beta +1)^{2}
\end{equation}
and is a confining potential for $\left\vert \alpha \right\vert ,\left\vert
\beta \right\vert >1/2$, i.e., in the region bounded by the singularities of $V(x;\alpha,\beta)$ at $x=0$ and $\pi/2$. Only in the case $|\alpha| = |\beta|$ the potential hole is symmetrical.\par
%
%
\sloppy

Consider a solution $\psi _{E}\left( x;\alpha ,\beta \right) $ of the Schr%
\"{o}dinger equation associated to this potential
\begin{equation}
  \left( \widehat{H}(x;\alpha ,\beta )-E\right) \psi _{E}\left( x;\alpha
  ,\beta \right) =0,
\end{equation}
where
\begin{equation}
  \widehat{H}(x;\alpha ,\beta )=-\frac{d^{2}}{dx^{2}}+V(x;\alpha ,\beta ).
\end{equation}
\par
%
%
Introducing the gauge transformation
\begin{equation}
  \psi _{E}\left( x;\alpha ,\beta \right) =\psi _{0}\left( x;\alpha ,\beta
  \right) y_{E}\left( x;\alpha ,\beta \right) ,
\end{equation}
with
\begin{equation}
  \psi _{0}\left( x;\alpha ,\beta \right) =\left( \sin x\right) ^{\alpha
  +1/2}\left( \cos x\right) ^{\beta +1/2},  \label{psi0}
\end{equation}
which is non zero on $\left] 0,\pi /2\right[ $ and satisfies the Schr\"odinger equation at ``zero energy''
\begin{equation}
  \widehat{H}(x;\alpha ,\beta )\psi _{0}\left( x;\alpha ,\beta \right) =0,
\end{equation}
we obtain for $y_{E}\left( x;\alpha ,\beta \right) $ the following second-order linear ODE
\begin{equation}
  y_{E}^{\prime \prime }\left( x;\alpha ,\beta \right) +2\left( \log \psi
  _{0}\left( x;\alpha ,\beta \right) \right) ^{\prime }y_{E}^{\prime }\left(
  x;\alpha ,\beta \right) +Ey_{E}\left( x;\alpha ,\beta \right) =0,
\end{equation}
where the prime denotes a derivative with respect to $x$.\par
%
%
Using the change of variable $z=\cos 2x$, $z\in \left] -1,1\right[ $, the
above equation becomes (with a dot denoting a derivative with respect to $z$)
\begin{equation}
  \left( 1-z^{2}\right) \ddot{y}_{E}\left( z;\alpha ,\beta \right) -
  \left[ \left( \alpha +\beta +2\right) z+\left( \alpha -\beta \right) \right] 
  \dot{y}_{E}\left( z;\alpha ,\beta \right) +\frac{E}{4}y_{E}\left(
  z;\alpha ,\beta \right) =0,  \label{hypergeo1}
\end{equation}
in which we recognize a hypergeometric equation \cite{erdelyi}.\par
%
%
Writing $E=E_{\nu }=4\nu \left( \nu +\alpha +\beta +1\right) $, we arrive at
\begin{equation}
\begin{split}
  & \left( 1-z^{2}\right) \ddot{y}_{\nu }\left( z;\alpha ,\beta \right) -
     \left[ \left( \alpha +\beta +2\right) z+\left( \alpha -\beta \right) \right] 
     \dot{y}_{\nu }\left( z;\alpha ,\beta \right)\\
  & \quad +\nu \left( \nu +\alpha +\beta +1\right) y_{\nu }\left( z;\alpha ,\beta \right) =0,  
\end{split}\label{eqjac}
\end{equation}
which, for integer values of the spectral parameter $\nu =n$ and values of
the $\alpha ,\beta $ parameters in the interval $\left] -1,+\infty \right[ $%
, admits the Jacobi polynomial \cite{szego,erdelyi} 
\begin{equation}
\begin{split}
  \mathit{P}_{n}^{\left( \alpha ,\beta \right) }\left( z\right) &=\frac{1}{%
    2^{n}}\sum_{k=0}^{n}\left( -1\right) ^{n-k}\binom{n+\alpha }{k}\binom{%
    n+\beta }{n-k}\left( 1-z\right) ^{n-k}\left( 1+z\right) ^{k} \\
  &=\frac{\Gamma \left( n+\alpha +1\right) }{n!\Gamma \left( n+\alpha +\beta
    +1\right) }\sum_{k=0}^{n}\left( -1\right) ^{k}\binom{n}{k}\frac{\Gamma
    \left( n+\alpha +\beta +1+k\right) }{2^{k}\Gamma \left( \alpha +1+k\right) }%
    \left( 1-z\right) ^{k}  \\
  &=\frac{\left( -1\right) ^{n}\Gamma \left( n+\beta +1\right) }{n!\Gamma
    \left( n+\alpha +\beta +1\right) }\sum_{k=0}^{n}\left( -1\right) ^{k}\binom{n%
    }{k}\frac{\Gamma \left( n+\alpha +\beta +1+k\right) }{2^{k}\Gamma \left(
    \beta +1+k\right) }\left( 1+z\right) ^{k}  
\end{split}
\end{equation}
as a solution analytic at the origin $z=0$.\par
%
%
Note that the standard Gauss form for the hypergeometric equation \cite%
{erdelyi} is obtained from Eq.~(\ref{hypergeo1}) via the second change of
variable $w=\left( 1-z\right) /2$, $w\in \left] 0,1\right[ $,
\begin{equation}
\begin{split}
  & \left\{w\left( 1-w\right) \frac{d^2}{dw^2} + \left[ \left( \alpha +1\right) -\left( \alpha +\beta +2\right) w
     \right] \frac{d}{dw} +\nu \left( \nu +\alpha +\beta +1\right)\right\} \\
  & \quad \times y_{\nu }\left( w;\alpha ,\beta \right) =0.
\end{split}
\end{equation}
%
%
{}From the preceding results, we deduce that, for $|\alpha| ,|\beta| >1/2$, the
physical spectrum of the TDPT potential associated to the asymptotic
Dirichlet boundary conditions
\begin{equation}
  \psi \left( 0^{+};\alpha ,\beta \right) =0=\psi \left( \left( \frac{\pi }{2}
  \right) ^{-};\alpha ,\beta \right)   \label{DBC}
\end{equation}
is given in terms of Jacobi polynomials $\mathit{P}_{n}^{\left( \alpha
,\beta \right) }$ by
\begin{equation}
  \left\{\begin{array}{c}
    \text{ }E_{n}\left( \alpha ,\beta \right) =(\alpha _{n}+\beta
       _{n}+1)^{2}-(\alpha +\beta +1)^{2}=4n(\alpha +\beta +1+n) \\ 
    \\ 
    \psi _{n}\left( x;\alpha ,\beta \right) =\psi _{0}\left( x;\alpha ,\beta
    \right) \mathit{P}_{n}^{\left( \alpha ,\beta \right) }\left( \cos 2x\right)
  \end{array}\right. ,\ 
  n\in \mathbb{N},
\end{equation}
with $(\alpha _{n},\beta _{n})=(\alpha +n,\beta +n)$.\par
%
%
The dispersion relation, $E_{\nu }$ as a function of $\nu $, is a convex
parabola with zeros at $\nu =0$ and $\nu =-\left( \alpha +\beta +1\right) $
and the disconjugacy sector corresponds to the values of $\nu $ in between
these two boundaries.\par
%
%
{}From the point of view of pure parameters transformations, we have
three possible discrete symmetries for $V(x;\alpha ,\beta )$, which are given
by

a)
\begin{equation}
  (\alpha ,\beta )\overset{\Gamma _{+}}{\longrightarrow }(-\alpha ,\beta
  ),\quad\left\{ 
    \begin{array}{c}
    V(x;\alpha ,\beta )\overset{\Gamma _{+}}{\longrightarrow }V(x;\alpha ,\beta
      )+4\alpha (\beta +1), \\ 
    \psi _{n}(x;\alpha ,\beta )\overset{\Gamma _{+}}{\longrightarrow }\phi
      _{n,+}(x;\alpha ,\beta )=\psi _{n}(x;-\alpha ,\beta ),%
    \end{array}
  \right.  \label{sym1}
\end{equation}

b)
\begin{equation}
  (\alpha ,\beta )\overset{\Gamma _{-}}{\longrightarrow }(\alpha ,-\beta
  ),\quad\left\{ 
    \begin{array}{c}
    V(x;\alpha ,\beta )\overset{\Gamma _{-}}{\longrightarrow }V(x;\alpha ,\beta
      )+4\beta (\alpha +1), \\ 
    \psi _{n}(x;\alpha ,\beta )\overset{\Gamma _{-}}{\longrightarrow }\phi
      _{n,-}(x;\alpha ,\beta )=\psi _{n}(x;\alpha ,-\beta ),%
    \end{array}
  \right.  \label{sym2}
\end{equation}

c)
\begin{equation}
  (\alpha ,\beta )\overset{\Gamma _{3}=\Gamma _{+}\circ \Gamma _{-}}{%
  \longrightarrow }(-\alpha ,-\beta ), \quad\left\{ 
    \begin{array}{c}
    V(x;\alpha ,\beta )\overset{\Gamma _{3}}{\longrightarrow }V(x;\alpha ,\beta
      )+4(\alpha +\beta ), \\ 
    \psi _{n}(x;\alpha ,\beta )\overset{\Gamma _{3}}{\longrightarrow }\phi
      _{n,3}(x;\alpha ,\beta )=\psi _{n}(x;-\alpha ,-\beta ).%
    \end{array}%
  \right.  \label{sym3}
\end{equation}
\par
%
%
In the $(\alpha ,\beta )$ plane, $\Gamma _{+}$ and $\Gamma _{-}$ correspond
respectively to the reflections with respect to the axes $\alpha =0$ and $%
\beta =0$. Combining the transformation of coordinates to the parameters
transformations, we have a supplementary symmetry given by
\begin{equation}
  (x;\alpha ,\beta )\overset{\Omega =S\otimes P}{\longrightarrow }(\xi ;\beta
  ,\alpha ), \quad\left\{ 
    \begin{array}{c}
    V(x;\alpha ,\beta )\overset{\Omega }{\longrightarrow }V(x;\alpha ,\beta ), \\ 
    \phi _{n,+}(x;\alpha ,\beta )\overset{\Omega}{\longrightarrow }\phi
      _{n,+}(\xi;\beta ,\alpha )= (-1)^n \phi _{n,-}(x;\alpha ,\beta ),%
    \end{array}%
  \right.
\end{equation}
where
\begin{equation}
  x\in \left[ 0,\pi /2\right] \overset{S}{\longrightarrow }\xi =\frac{\pi }{2}%
  -x\in \left[ 0,\pi /2\right]
\end{equation}
is the reflection on the coordinate axis with respect to the point $\frac{%
\pi }{4}$ ($\frac{\pi }{4}-x\overset{S}{\longrightarrow }x-\frac{\pi }{4}$)
and
\begin{equation}
  (\alpha ,\beta )\overset{P}{\longrightarrow }(\beta ,\alpha )  \label{symsup}
\end{equation}
is the reflection with respect to the principal diagonal in the $(\alpha
,\beta )$ parameters plane.\par
%
%
The functions $\phi _{n,i},\ i=+,-,3$, satisfy the respective equations
\begin{equation}
  \widehat{H}\left( \alpha ,\beta \right) \phi _{n,i}(x;\alpha ,\beta )=%
  \mathcal{E}_{n,i}(\alpha ,\beta )\phi _{n,i}(x;\alpha ,\beta )\ ,
\label{EdSregL1}
\end{equation}
with
\begin{equation}
  \left\{ 
    \begin{array}{c}
    \mathcal{E}_{n,+}(\alpha ,\beta )=E_{n-\alpha }(\alpha ,\beta )=4(n-\alpha
      )(n+\beta +1), \\ 
    \mathcal{E}_{n,-}(\alpha ,\beta )=E_{n-\beta }\left( \alpha ,\beta \right) =%
      \mathcal{E}_{n,+}(\beta ,\alpha ), \\ 
     \mathcal{E}_{n,3}(\alpha ,\beta )=E_{-\left( n+1\right) }(\alpha ,\beta ) =
       - 4 (n+1) (\alpha + \beta - n).
     \end{array}%
  \right. \   \label{Energiesneg}
\end{equation}
\par
%
%
The eigenvalue $\mathcal{E}_{n,3}$ is negative for $n<\alpha +\beta $, while
for $i=+$ and $i=-$, the disconjugacy condition $\mathcal{E}_{n,i}(\alpha ,\beta
)\leq 0$ imposes that the constraints $\alpha >n$ and $\beta >n$ be satisfied, respectively.\par
%
%
\section{DISCONJUGATED SEED FUNCTIONS ASSOCIATED TO PARA-JACOBI POLYNOMIALS}
\setcounter{equation}{0}

Suppose that $\alpha $ and $\beta $ are two positive integers
\begin{equation}
  \alpha =N\in 
  \mathbb{N}
  ^{\ast },\quad \beta =M\in 
  \mathbb{N}
  ^{\ast },  \label{alphbet}
\end{equation}
and apply the $\Gamma _{3}$ symmetry. The function $\phi _{n,3}(x;N,M)=\psi _{n}(x;-N,-M)$
is a solution of the Schr\"{o}dinger equation
\begin{equation}
  \left( -\frac{d^{2}}{dx^{2}}+V(x;N,M)-\mathcal{E}_{n,3}(N,M)\right) \phi
  _{n,3}(x;N,M)=0,  \label{Eds4}
\end{equation}
where
\begin{equation}
  \mathcal{E}_{n,3}(N,M)=E_{-\left( n+1\right) }(N,M)<0,
\end{equation}
if $n<N+M$.\par
%
%
But $\phi _{n,3}(x;N,M)=\psi _{n}(x;-N,-M)$ can also be written as
\begin{equation}
  \phi _{n,3}(x;N,M)=\psi _{0}\left( z;-N,-M\right) y_{n}\left( z;-N,-M\right),
\end{equation}
where, up to a constant factor,
\begin{equation}
  \psi _{0}\left( z;-N,-M\right) =\left( 1-z\right) ^{\left( -N+1/2\right)
  /2}\left( 1+z\right) ^{\left( -M+1/2\right) /2}
\end{equation}
and
\begin{equation}
\begin{split}
  & \left( 1-z^{2}\right) \ddot{y}_{n}\left( z;-N,-M\right) -\left[
      \left( -N-M+2\right) z+\left( -N+M\right) \right] \dot{y}_{n}\left(z;-N,-M\right) \\
  & \quad +n\left( n-N-M+1\right) y_{n}\left( z;-N,-M\right) =0.
\end{split}
\end{equation}
\par
%
%
In this case, as noticed by Szeg\"o \cite{szego} and emphasized by Calogero
and Yi \cite{calogero}, for values of $n$ such that
\begin{equation}
  \frac{N+M}{2}\leq n<N+M,  \label{szegocond}
\end{equation}
the general solution of the equation for $y_{n}\left( z;-N,-M\right)$ is a
polynomial, called \textbf{para-Jacobi polynomial}, which has the (monic)
form \cite{calogero}
\begin{equation}
\begin{split}
  p_{n}^{\left( -N,-M\right) }\left( z;\lambda \right) &=\frac{\left(
    -2\right) ^{n}(n-M)!n!}{(2n-M-N)!}\sum_{k=0}^{n-M}\frac{\left( -1\right)
    ^{n-k}\left( 2n-M-N-k\right) !}{k!(n-M-k)!(n-k)!}\left( \frac{1+z}{2}\right)
    ^{n-k} \\
  &\quad {}+\lambda \frac{\left( -2\right) ^{n}(2n-M-N+1)!(M+N-n-1)!}{(n-N)!} \\
  &\quad {} \times\sum_{k=2n-M-N+1}^{n}\frac{\left( -1\right) ^{n-k}\left( k-n+M-1\right) !}{
    k!(k+N+M-2n-1)!(n-k)!}\left( \frac{1+z}{2}\right) ^{n-k},  
\end{split} \label{parajac}
\end{equation}
$\lambda $ being an arbitrary real parameter.\par
%
%
In other words, we have a basis of two polynomial eigenfunctions
\begin{equation}
  \left\{ 
    \begin{array}{c}
    \Theta _{n,1}^{\left( -N,-M\right) }\left( z\right) =\sum_{k=M}^{n}%
      \frac{\left( -1\right) ^{k}\left( n-M-N+k\right) !}{2^{k}k!(k-M)!(n-k)!}%
      \left( 1+z\right) ^{k}, \\ 
    \Theta _{n,2}^{\left( -N,-M\right) }\left( z\right)
      =\sum_{k=0}^{N+M-n-1}\frac{\left( -1\right) ^{k}\left( M-1-k\right) !%
      }{2^{k}k!(N+M-n-1-k)!(n-k)!}\left( 1+z\right) ^{k},%
    \end{array}%
  \right.  \label{theta}
\end{equation}
with $N,M>0$ and
\begin{equation}
  \left\{ 
    \begin{array}{c}
    \Theta _{n,1}^{\left( -N,-M\right) }\left( -1\right) =0, \\ 
    \Theta _{n,2}^{\left( -N,-M\right) }\left( -1\right) =\frac{\left(
      M-1\right) !}{(N+M-n-1)!n!}>0.%
    \end{array}%
  \right.
\end{equation}
\par
%
%
It results that the general solution of Eq.~(\ref{Eds4}) can be
written as ($z=\cos 2x$)
\begin{equation}
  \psi _{n}\left( x;-N,-M;\lambda \right) =\psi _{0}\left( x;-N,-M\right)
  p_{n}^{\left( -N,-M\right) }\left( z;\lambda \right) ,  \label{psipara}
\end{equation}
with
\begin{equation}
\begin{split}
  p_{n}^{\left( -N,-M\right) }\left( z;\lambda \right) &=\frac{\left(
    -2\right) ^{n}(n-M)!n!}{(2n-M-N)!}\Theta _{n,1}^{\left( -N,-M\right) }\left(
    z\right) \\
  &\quad +\lambda \frac{\left( -2\right) ^{n}(2n-M-N+1)!(M+N-n-1)!}{(n-N)!}\Theta
    _{n,2}^{\left( -N,-M\right) }\left( z\right).  
\end{split} \label{parajac2}
\end{equation}\par
%
%
Moreover, from Eq.~(\ref{theta}), we have
\begin{equation}
\begin{split}
  \dot{\Theta }_{n,1}^{\left( -N,-M\right) }\left( z\right)&=
    \sum_{j=M-1}^{n-1}\frac{\left( -1\right) ^{j+1}\left(
    n-M-N+j+1\right) !}{2^{j+1}j!(j+1-M)!(n-j-1)!}\left( 1+z\right) ^{j} \\
  &=-\frac{1}{2}\Theta _{n-1,1}^{\left( -\left( N-1\right) ,-\left( M-1\right) \right)
    }\left( z\right).
\end{split}
\end{equation}
More generally, for $i=1,2$,
\begin{equation}
  \dot{\Theta }_{n,i}^{\left( -N,-M\right) }\left( z\right) =
  -\frac{1}{2}\Theta _{n-1,i}^{\left( -N+1,-M+1\right) }\left( z\right),  \label{derivtheta}
\end{equation}
which gives
\begin{equation}
  \dot{p}_{n}^{\left( -N,-M\right) }\left( z;\lambda \right)
  =n p_{n-1}^{\left( -N+1,-M+1\right) }\left( z;\lambda ^{\prime }\right),  \label{derivp}
\end{equation}
where
\begin{equation}
  \lambda ^{\prime }=\frac{M+N-n-1}{n}\lambda.  \label{lambdaprim}
\end{equation}
%
%
{}For $\frac{N+M}{2}\leq n<N+M$, $\psi _{n}\left( x;-N,-M;\lambda \right) $ is
disconjugated on $z\in \left] -1,1\right[ $ and consequently admits at most
one zero on this interval. To ensure the absence of node, we have to verify
that the sign of $\psi _{n}\left( x;-N,-M;\lambda \right) $ at the
boundaries of the interval is the same. This is the case if
\begin{equation}
  {\rm sign}\left( p_{n}^{\left( -N,-M\right) }\left( 1;\lambda \right) \right)
  = {\rm sign}\left( p_{n}^{\left( -N,-M\right) }\left( -1;\lambda \right) \right).
\end{equation}
\par
%
%
But due to the symmetry given in Eq.~(\ref{symsup}), we have \cite{calogero}
\begin{equation}
  p_{n}^{\left( -N,-M\right) }\left( -z;\lambda \right) =\left( -1\right)
  ^{n}p_{n}^{\left( -M,-N\right) }\left( z;\widetilde{\lambda }\right) ,
\end{equation}
where
\begin{equation}
\begin{split}
  \widetilde{\lambda } &=\left( -1\right) ^{2n-N-M+1}\lambda \\
  &\quad +\left( -1\right)^{n-M}\frac{n!\left( n-M\right) !\left( n-N\right) !}{\left( 2n-N-M\right)
    !\left( 2n-N-M+1\right) !(N+M-n-1)!}.
\end{split}
\end{equation}
\par
%
%
This implies
\begin{equation}
  p_{n}^{\left( -N,-M\right) }\left( -1;\lambda \right) =\lambda \frac{\left(
  -2\right) ^{n}\left( 2n-N-M+1\right) !\left( M-1\right) !}{(n-N)!\,n!}
\end{equation}
and
\begin{equation}
\begin{split}
  p_{n}^{\left( -N,-M\right) }\left( 1;\lambda \right) &=\left( -1\right)
    ^{n}p_{n}^{\left( -M,-N\right) }\left( -1;\widetilde{\lambda }\right)\\
  &=\left( -1\right) ^{n}\widetilde{\lambda }\frac{\left( -2\right) ^{n}\left(
    2n-N-M+1\right) !\left( M-1\right) !}{(n-N)!\,n!}.
\end{split}
\end{equation}
\par
%
%
The absence of node is then obtained if $\lambda $ and $\left( -1\right) ^{n}
\widetilde{\lambda }$ have the same sign. In the following we note that
\begin{equation}
  \lambda _{n}^{(-N,-M)}\equiv \frac{n!\left( n-M\right) !\left( n-N\right) !}{%
  \left( 2n-N-M\right) !\left( 2n-N-M+1\right) !(N+M-n-1)!}>0.
\end{equation}
\par
%
%
\noindent
\textit {Case (i)}

Suppose first that we take $\lambda >0$. We must then have
\begin{equation}
  \left( -1\right) ^{n}\widetilde{\lambda }=\left( -1\right) ^{n-N-M+1}\lambda
  +\left( -1\right) ^{M}\lambda _{n}^{(-N,-M)}>0,
\end{equation}
that is, if $n-N-M$ is odd,
\begin{equation}
  \lambda >\left( -1\right) ^{M+1}\lambda _{n}^{(-N,-M)}.
\end{equation}
This is always achieved when $M$ is even ($n-N$ is odd) and it necessitates
to take
\begin{equation}
  \lambda >\lambda _{n}^{(-N,-M)}
\end{equation}
when $M$ is odd ($n-N$ is even). In the case where $n-N-M$ is even, we deduce in the same way that the
condition of equality of signs can never be reached if $M$ is odd ($n-N$ is
odd) and imposes to take
\begin{equation}
  0<\lambda <\lambda _{n}^{(-N,-M)}
\end{equation}
when $M$ is even ($n-N$ is even).\par
%
%
\noindent
\textit {Case (ii)}

If we take $\lambda <0$, we must have
\begin{equation}
  \left( -1\right) ^{n}\widetilde{\lambda }=\left( -1\right) ^{n-N-M+1}\lambda
  +\left( -1\right) ^{M}\lambda _{n}^{(-N,-M)}<0,
\end{equation}
that is, if $n-N-M$ is odd,
\begin{equation}
  \lambda <\left( -1\right) ^{M+1}\lambda _{n}^{(-N,-M)}.
\end{equation}
This is always achieved when $M$ is odd ($n-N$ is even) and it necessitates
to take
\begin{equation}
  \left\vert \lambda \right\vert >\lambda _{n}^{(-N,-M)}
\end{equation}
when $M$ is even ($n-N$ is odd). In the case where $n-N-M$ is even, we deduce in the same way that the
condition of equality of signs can never be reached if $M$ is even ($n-N$ is
even) and imposes to take
\begin{equation}
  \left\vert \lambda \right\vert <\lambda _{n}^{(-N,-M)}
\end{equation}
when $M$ is odd ($n-N$ is odd).\par
%
%
To summarize, $\psi _{n}\left( x;-N,-M;\lambda \right) $ has no node in the four
following cases:
\begin{align*}
  \text{(i)} \; & \text{$M$ is even, $n-N$ is odd, $\lambda <-\lambda _{n}^{(-N,-M)}$ or $\lambda>0$;} \\
  \text{(ii)} \; & \text{$M$ is even, $n-N$ is even, $0<\lambda <\lambda _{n}^{(-N,-M)}$;} \\
  \text{(iii)} \; & \text{$M$ is odd, $n-N$ is even, $\lambda <0$ or $\lambda >\lambda_{n}^{(-N,-M)}$;} \\
  \text{(iv)} \; & \text{$M$ is odd, $n-N$ is odd, $-\lambda _{n}^{(-N,-M)}<\lambda <0$.} 
\end{align*}
\par
%
%
\section{REGULAR RATIONAL EXTENSIONS}
\setcounter{equation}{0}

Suppose $\lambda $ satisfies the conditions mentioned above. Then $\psi
_{n}\left( x;-N,-M;\lambda \right) $ can be used as seed function to build a
state-adding DBT, which generates a regular rational extension of $%
V(x;N-1,M-1)$. It is given by (see Eqs.~(\ref{pottrans}) and (\ref{psipara}))
\begin{equation}
\begin{split}
  \widetilde{V}^{\left( n\right) }(x;N,M;\lambda ) &= V(x;N,M)-2\left( \log
     \left( \phi _{n,3}\left( x;N,M\right) \right) \right) ^{\prime \prime } \\
  &= V(x;N,M)-2\left( \log \left( \psi _{n}\left( x;-N,-M;\lambda \right)
     \right) \right) ^{\prime \prime } \\
  &= V(x;N-1,M-1)-4(N+M)-2\left( \log \left( p_{n}^{\left( -N,-M\right)
     }\left( z;\lambda \right) \right) \right) ^{\prime \prime }, 
\end{split}
\end{equation}
since
\begin{equation}
\begin{split}
  \widetilde{V}^{\left( 0\right) }(x;N,M) &= V(x;N,M)-2\left( \log \left( \psi
      _{0}\left( x;-N,-M\right) \right) \right)^{\prime \prime } \\
  &= V(x;N-1,M-1)+E_{-1}\left( N,M\right) . 
\end{split}
\end{equation}
\par
%
%
On taking into account of the relations
\begin{equation}
 \frac{d}{dx}=-2\sqrt{1-z^{2}}\frac{d}{dz},\qquad \frac{d^{2}}{dx^{2}}=4\left(
  1-z^{2}\right) \frac{d^{2}}{dz^{2}}-4z\frac{d}{dz},
\end{equation}
and of Eqs. (\ref{derivp}) and (\ref{lambdaprim}), we arrive at
\begin{equation}
\begin{split}
  & \widetilde{V}^{\left( n\right) }(x;N,M;\lambda ) \\ 
  & \quad = V(x;N-1,M-1)-4(N+M) 
      -8n(n-1)\left( 1-z^{2}\right) \frac{p_{n-2}^{\left(-N+2,-M+2\right) }
      \left( z;\lambda ^{\prime \prime }\right) }{p_{n}^{\left(-N,-M\right) }\left( z;\lambda \right) } \\
  & \qquad {} +8n^{2}\left( 1-z^{2}\right) \left( \frac{p_{n-1}^{\left( -N+1,-M+1\right)
      }\left( z;\lambda ^{\prime }\right) }{p_{n}^{\left( -N,-M\right) }\left(
      z;\lambda \right) }\right) ^{2}+8nz\frac{p_{n-1}^{\left( -N+1,-M+1\right)
      }\left( z;\lambda ^{\prime }\right) }{p_{n}^{\left( -N,-M\right) }\left(
      z;\lambda \right) }, 
\end{split}
\end{equation}
where
\begin{equation}
  \lambda ^{\prime \prime }=\frac{\left( M+N-n-1\right) \left( M+N-n-2\right) 
  }{n(n-1)}\lambda .
\end{equation}
\par
%
%
{}For $\widetilde{V}^{\left( n\right) }(x;N,M;\lambda )$ to be a
confining potential on $\left] 0,\pi /2\right[ $, we need to impose that $%
V(x;N-1,M-1)$ has this property, which is achieved for $N,M>3/2$, thence $%
N,M\ge 2$. For such $N,M$ values, we note that $\widetilde{V}^{\left( n\right)
}(x;N,M;\lambda )$ is actually strongly repulsive in both $0$ and $\pi /2$,
since the singularities are there of the type $g/x^{2}$ ($g\geq 3/4$) and $%
g/\left( \pi /2-x\right) ^{2}$ ($g\geq 3/4$), respectively. This means that
at each extremity, only one basis solution is quadratically integrable. This
choice limits (see Eq.(\ref{szegocond})) the possible values of $n$ to $n\geq
2$, but in this case the corresponding Hamiltonian $\widehat{\widetilde{H}}%
^{\left( n\right) }(x;N,M;\lambda )$ is essentially self-adjoint \cite%
{frank,lathouwers}. The simplest example corresponds to $n=N=M=2$, in which case
\begin{equation}
  p_{2}^{\left( -2,-2\right) }\left( z;\lambda \right) =z^{2}+2(1-\lambda )z+1
\end{equation}
with $0<\lambda <\lambda _{2}^{(-2,-2)}=2$, which is example (6b) in Ref.~\cite{calogero}. 
The initial potential is 
\begin{equation}
  V(x;2,2)=\frac{15}{1-z^{2}}-25
\end{equation}
and the partner potential reads
\begin{equation}
\begin{split}
  \widetilde{V}^{\left( 2\right) }(x;2,2;\lambda ) &= \frac{3}{1-z^{2}}-16
     \frac{\left( \lambda -1\right) z+2\lambda ^{2}-4\lambda -1}{z^{2}+2(1-\lambda )z+1} \\
  &\quad {}+64\lambda (\lambda -2)\frac{(1-\lambda )z+1}{\left[ z^{2}+2(1-\lambda
     )z+1\right)] ^{2}}-25.
\end{split}
\end{equation}
The ground-state energy of the latter is equal to $E_{-3}(2,2)=-24$. In Fig.~1, the
corresponding potentials are plotted for different values of the $\lambda $ parameter.\par
%
%
\begin{figure}[h]
\begin{center}
\includegraphics{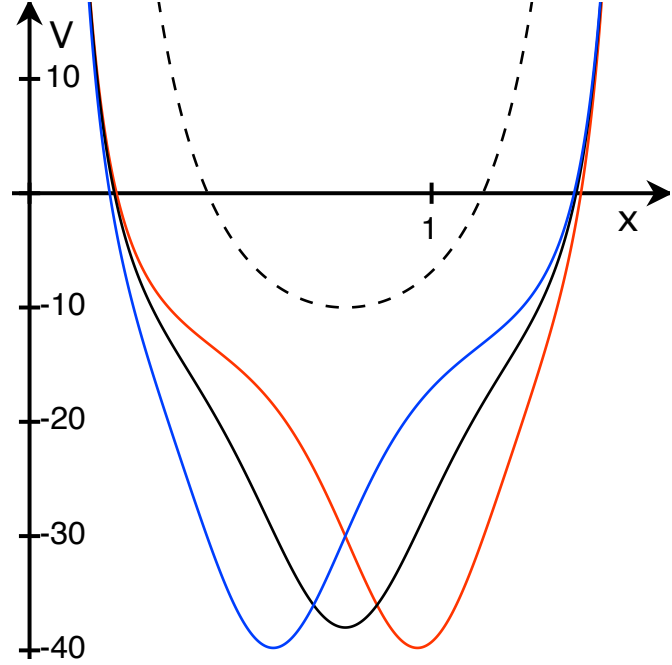}
\caption{Plot of $\widetilde{V}^{\left( 2\right) }(x;2,2;\lambda )$ as a function of $x$, for $\lambda=1/2$ (red line), $\lambda=1$ (black solid line), and $\lambda=3/2$ (blue line). The partner $V(x;2,2)$ is also shown (black dashed line).}
\end{center}
\end{figure}
\par
%
%
Using the standard properties of Wronskians \cite{muir}
\begin{equation}
  \left\{ 
    \begin{array}{c}
    W\left( uy_{1},...,uy_{m}\mid x\right) =u\left( x\right) ^{m}W\left(
      y_{1},...,y_{m}\mid x\right),  \\ 
    W\left( y_{1},...,y_{m}\mid x\right) =\left( \frac{dz}{dx}\right)
      ^{m(m-1)/2}W\left( y_{1},...,y_{m}\mid z\right) ,
    \end{array}
  \right. 
\end{equation}
the eigenstates of $\widetilde{V}^{\left( n\right) }(x;N,M;\lambda )$ are
given by ($k\geq 0$)
\begin{equation}
\begin{split}
  \widetilde{\psi }_{k}^{\left( n\right) }(x;N,M;\lambda ) &= \frac{W(\psi_{n}\left( x;-N,-M;\lambda \right) ,\psi    
     _{k}\left( x;N,M\right) \mid x)}{\psi _{n}\left( x;-N,-M;\lambda \right) } \\
  &\propto \psi _{0}\left( x;-N,-M\right) \left( 1-z\right) ^{1/2}\left(1+z\right) ^{1/2} \\
  & \quad {}\times \frac{W\bigl(p_{n}^{\left( -N,-M\right) }\left( z;\lambda
     \right) ,\left( 1-z\right) ^{N}\left( 1+z\right) ^{M}\mathit{P}_{k}^{\left(
     N,M\right) }\left( z\bigr) \mid z\right)}{p_{n}^{\left( -N,-M\right) }\left(
     z;\lambda \right) },  
\end{split}
\end{equation}
that is, on using derivation properties of Jacobi polynomials \cite%
{szego,erdelyi} and Eq.~(\ref{derivp})
\begin{equation}
  \widetilde{\psi }_{k}^{\left( n\right) }(x;N,M;\lambda )\propto \frac{\psi
  _{0}\left( x;N-1,M-1\right) }{p_{n}^{\left( -N,-M\right) }\left( z;\lambda
  \right) }Q_{k}^{\left( n\right) }\left( z;N,M;\lambda \right) ,
\end{equation}
where
\begin{equation}
\begin{split}
  Q_{k}^{\left( n\right) }\left( z;N,M;\lambda \right)  &=\left(
    1-z^{2}\right) \biggl( \frac{k+M+N+1}{2}\mathit{P}_{k-1}^{\left(
    N+1,M+1\right) }\left( z\right) p_{n}^{\left( -N,-M\right) }\left( z;\lambda
    \right) \\
  &\quad {}-np_{n-1}^{\left( -N+1,-M+1\right) }\left( z;\lambda ^{\prime
    }\right) \mathit{P}_{k}^{\left( N,M\right) }\left( z\right) \biggr) \\
  &\quad {}-\left( \left( N+M\right) z+N-M\right) \mathit{P}_{k}^{\left( N,M\right)
    }\left( z\right) p_{n}^{\left( -N,-M\right) }\left( z;\lambda \right) . 
\end{split} \label{Q}
\end{equation}
\par
%
%
Moreover, since
\begin{equation}
  \widetilde{\psi }_{-(n+1)}^{\left( n\right) }(x;N,M;\lambda )=\frac{1}{\psi
  _{n}\left( x;-N,-M;\lambda \right)} =\frac{\psi _{0}\left( x;N-1,M-1\right) }{
  p_{n}^{\left( -N,-M\right) }\left( z;\lambda \right) }
\end{equation}
satisfies the Dirichlet boundary conditions and the square integrability
one, it is an eigenstate of $\widetilde{V}^{\left( n\right)
}(x;N,M;\lambda )$ with an energy $E_{-(n+1)}(N,M)$ and the DBT $\widehat{A}\left(
w _{n,3}\right) $\ is state-adding. To summarize, the spectrum of the
extended potential is ($k\in \left\{ -(n+1),0,1,...\right\} $)
\begin{equation}
  \left\{ 
    \begin{array}{c}
    E_{k}(N,M), \\[0.2cm] 
    \widetilde{\psi }_{k}^{\left( n\right) }(x;N,M;\lambda )\propto \frac{\psi
      _{0}\left( x;N-1,M-1\right) }{p_{n}^{\left( -N,-M\right) }\left( z;\lambda
      \right) }Q_{k}^{\left( n\right) }\left( z;N,M;\lambda \right) ,
    \end{array}
  \right. \quad 
\end{equation}
where $Q_{-n-1}^{\left( n\right) }\left( z;N,M;\lambda \right) =1$ and where
the $Q_{k\geq 0}^{\left( n\right) }$ are given in Eq.~(\ref{Q}).\par
%
%
Due to the orthogonality properties of the $\widetilde{\psi }_{k}^{\left(
n\right) }$, we deduce that the $Q_{k}^{\left( n\right) }\left(
z;N,M;\lambda \right) $ are orthogonal polynomials on $\left] -1,1\right[ $
with respect to the measure
\begin{equation}
\mu _{n}^{\left( -N,-M\right) }\left( z;\lambda \right) =\frac{%
(1-z)^{N-1}(1+z)^{M-1}}{\left( p_{n}^{\left( -N,-M\right) }\left( z;\lambda
\right) \right) ^{2}}.  \label{mu}
\end{equation}
\par
%
%
\section{CONCLUSION}

In this article we have shown that it is possible to build one-step regular
rational extensions of the TDPT potential depending both on a integer index $%
n$ and on a continuously varying parameter $\lambda $. This is achieved by
using, for the underlying Darboux transformations, seed functions which are
associated to the para-Jacobi polynomials of Calogero and Yi . To each value
of $n$ corresponds then a novel family of $\lambda $-dependent polynomials
which are orthogonal on $\left] -1,1\right[ $.\par
%
%
Note that some continuously parametrized rationally extended potentials can
also be obtained by using an extension scheme based on confluent DBT \cite%
{GQ1}.\par
%
%
The multistep version of the results presented here is in progress and will
be the subject of a forthcoming paper.
%
%
\section*{ACKNOWLEDGMENTS}

For one of us (B.B.) it is a great pleasure to thank Yves Grandati as well
as the other members of BioPhysStat group at Universit\'{e} de Lorraine-Metz
for their warm hospitality. The remaining two of us (Y.G.\ and C.Q.) would like to thank F.\ Calogero for some very interesting discussions.\par
%
%

\end{document}